# Intrinsic Geometric Structure of Turbulent Flow for Newton Fluid

*Xiao Jianhua*
*Shanghai Jiao Tong University, Natural Science Foundation Research Group*

**Abstract:** Many researches show that the complicated motion of fluid, such as turbulence, cannot be well solved by the Navier-Stokes equation. Some researches even go to doubt its effectiveness for turbulence. Chen Zida has founded that the definition of vortex, based on the Stokes's decomposition, cannot well describe the local rotation when the velocity gradient is highly asymmetric. Chen reformulates the Stokes's S+R decomposition into a general S+R decomposition. By further extending Chen's results, this research studies the motion equation of fluid for the case where highly asymmetric velocity gradient is exhibited. The result shows that the classical NS equation does not meet the requirement of angular momentum conservation, which is apparently ignored for infinitesimal velocity gradient of fluid. This paper reformulates the intrinsic geometric description of fluid motion and two additional equations are introduced. Combining with the classical NS equation, the reformulated motion equations are in closed-form. The research shows that the NS equation is good approximation for average flow, so it can not solve the turbulent problem in essential sense. However, this conclusion does not deny that with suitable additional condition for special engineering problem it is still a would-be acceptable approximation. The research concludes that: (a) large velocity gradient flow (no matter it is steady or un-steady) will produce significant non-symmetric stress; (b) un-steady flow (no matter it is infinitesimal or not) will produce non-symmetric stress. The non-symmetric stress produces local rotation. When the local rotation is big enough, the flow will become turbulent flow. The research shows that there are three ways to determine the turbulent transition condition: (1) solving fluid motion equation; (2) using the intrinsic parameters of fluid material; (3) using pressure condition with given material parameters. Therefore, the essential feature of turbulent flow is well described in this research. Related equations are given with brief description.

## 1. Introduction

The Navier-Stokes equations are widely used in fluid mechanics as basic equations. Its effectiveness for simple fluid motion is evident. In fact, NS equation has been unqualifiedly used as the unique foundation of whole fluid dynamics. But, recently, enquiry about its reasonability and solvability has been proposed by many researchers [1-2]. Some even goes to claim that the turbulence cannot be solved by the NS equation [3].

This problem in essential sense has been attacked by Lodge A S (1974). He claimed that except to use body tensor (defined on material under consideration) to establish motion equations the fluid motion cannot be correctly described [4]. His concept of material point is well accepted rather his idea of body tensor, which is difficult to understand.

Chen Zhida's research shows that, introducing commoving dragging coordinator system, the deformation gradient, in fact, is base vectors transformation among initial configuration and current configuration of continuum. In fluids, the deformation gradient on commoving dragging coordinator system can be defined by the inertial velocity gradient. By this way, Lodge's body tensor concept can be achieved.

The most important advancement for the concept is that Chen's research shows that the transformation can be decomposed into the addition of one symmetric tensor expressing stretching and one unit orthogonal tensor expressing local rotation [5]. This feature makes that the fluid motion can be decomposed into a non-rotational flow and a rotational flow. Mathematically, in classical fluid mechanics, the velocity gradient can be divided into symmetry part and asymmetry part, where Stokes gave rotation interpretation for infinitesimal asymmetry part. Hence, it is clear the rotational flow is related with the asymmetry part of velocity gradient. In fact, based on Chen's research, the asymmetry part is related with an orthogonal rotation. More important point is that Chen shows that the symmetry part of transformation contains the contribution of orthogonal rotation. Therefore, geometrically, rotation and stretching have intrinsic relationship rather than be independent. Chen shows that the definition of vortex, based on the Stokes' S+R decomposition [5], cannot well describe the local rotation which is essentially different from the global rotation of fluid, when the average local rotation is the main role. However, for small average local rotation, the definition of vortex is a good approximation for local rotation. Hence, one can only say that the vortex is ill-defined. The ill-definition of vortex is the main cause which misleads many researchers on rotational flows. As the consequence of above mentioned problems, the NS equation is ill-used in fluid mechanics.

Chen's S-R decomposition can be extended to large deformation and large rotation cases with the means of discarding the unit-orthogonal rotation and replacing it with an orthogonal rotation with expanding [6]. So, geometrically, Chen's research forms the basis for a complete fluid motion geometry.

Once the orthogonal rotation tensor is introduced, the stress tensor will be non-symmetric. In classical mechanics, the non-symmetric stress is directly contradictive with the NS equation, as the uniqueness of NS equation is destroyed. This problem has discouraged many potential researchers.

Fortunately, Chen's rational mechanics theory forms basis for further research on this sharp problem. This research shows that, the traditional equilibrium equation (NS equation) only meets moment conservation, and does not meet the local angular-momentum conservation. The angular-moment conservation shown in the paper does accept the non-symmetry stress related with rotation. It is confirmed that for infinitesimal non-rotational flow the NS equation indeed is accurate. However, for any large flow, the stress must be non-symmetric. Now, it is clear that infinitesimal flow and large flow have intrinsic difference.

Based on Chen's theory, the symmetry part of motion transformation (be defined as strain rate in Chen's Theory) contains a coupling tensor directly related with rotation tensor, this coupling tensor makes that the strain rate defined traditionally significantly differ from the actual value calculated based on measured stress field. The additional strain rate related with the coupling tensor is the function of square of elements of asymmetry part of velocity gradient. Traditionally, this item is introduced by an additional stress, such as Reynolds stress. Based on this research, the classical NS equation is correct but not complete for complicated flow and for turbulence. After combining with two new equations related with local angular momentum conservation, the dynamic equations of fluid motion are completely established.

This paper will firstly introduce the related geometrical tensors and then formulate the motion equations for fluid motion. After that, two typical rotation motion equations are discussed. One is for incompressible rotation flow which not only has no volume variation but also has no linear momentum variation. When the local rotation angular meets some conditions, the rotational flow will transit into turbulent flow. The research gives the geometrical condition and the turbulent pressure condition for turbulent transition. Another kind of rotation is compressible rotational flow which always has velocity increasing. For isothermal flow, the velocity increasing gives the bubble volume density. The geometrical condition and the turbulent pressure condition for turbulent transition and instability are given by related equations. Finally, the intrinsic geometric structure of turbulent flow and the critical condition for turbulent transition are summarized.

## 2. Intrinsic Geometric Description of Fluid Motion

For a selected Lagrange coordinator system $(x^1, x^2, x^3)$ on the continuum, it is a Riemann three-dimension space. Chen Zhida defines the commoving dragging coordinator system as such a Lagrange coordinator system that its base vectors transforms according to the deformation of continuum while the Lagrange coordinator parameterization maintain the same for initial configuration and current configuration. For a material point $x^i$, there is a finite unit volume fluid element covering it. If the number of particles within the unit volume covering is enough to present a significant macroscopic representation, it is reasonable to introduce macroscopic representation of the inertial velocity $u^i$ and density $\rho$ defined at the material point

The velocity transformation tensor can be expressed by the gradient of velocity $u^i$ as following:

$$F_j^i = u^i\big|_j + \delta_j^i \qquad (1)$$

where, $u^i\big|_j$ express the covariant differentiate of velocity fields; $\delta_j^i$ is Kronecker delta. For the commoving dragging coordinator system, its base vectors is transformed by the transformation $F_j^i$ from initial configuration to current configuration (for detailed discussion, please refer Appendix A). Note that the whole motion of fluid (no matter it is rotational or non-rotational) has zero velocity gradient on the commoving dragging coordinator system, hence, the velocity gradient represents the local relative motion, which is named as deformation in solid continuum.

Chen Zhida has shown that the velocity transformation can be decomposed into the addition of one symmetry tensor expressing stretching and one unit orthogonal tensor expressing local rotation [2-5]. That is, we have:

$$F_j^i = S_j^i + R_j^i \qquad (2)$$

Where:

$$S_j^i = \frac{1}{2}(u^i\big|_j + u^i\big|_j^T) - (1-\cos\Theta)L_k^i L_j^k \qquad (3)$$

$$R_j^i = \delta_j^i + \sin\Theta \cdot L_j^i + (1-\cos\Theta)L_k^i L_j^k \qquad (4)$$

$$L_j^i = \frac{1}{2\sin\Theta}(u^i\big|_j - u^i\big|_j^T) \qquad (5)$$

$$\sin\Theta = \frac{1}{2}[(u^1\big|_2 - u^2\big|_1)^2 + (u^2\big|_3 - u^3\big|_2)^2 + (u^3\big|_1 - u^1\big|_3)^2]^{\frac{1}{2}} \qquad (6)$$

In above expressions, T represents transpose, the parameter $\Theta$ is called local average rotation angular and tensor $L_j^k$ defines the local average rotation axis direction. Here, the average angular range is $(-\frac{\pi}{2}, \frac{\pi}{2})$.

But, the definition of local rotation angular is too strong to be reasonable for large shear deformation, as it requires the condition of:

$$\frac{1}{2}[(u^1|_2 - u^2|_1)^2 + (u^2|_3 - u^3|_2)^2 + (u^3|_1 - u^1|_3)^2]^{\frac{1}{2}} \leq 1 \tag{7}$$

This problem can be overcome by defining the local average rotation angular $\theta$ as the following:

$$(\cos\theta)^{-2} = 1 + \frac{1}{4}[(u^1|_2 - u^2|_1)^2 + (u^2|_3 - u^3|_2)^2 + (u^3|_1 - u^1|_3)^2] \tag{9}$$

It is easy to verify that Stokes-Chen's S-R additive decomposition theorem can be extended as:

$$\tilde{F}_j^i = \tilde{S}_j^i + (\cos\theta)^{-1}\tilde{R}_j^i \tag{10}$$

Where:

$$\tilde{S}_j^i = \frac{1}{2}(u^i|_j + u^i|_j^T) - (\frac{1}{\cos\theta} - 1)(\tilde{L}_k^i \tilde{L}_j^k + \delta_j^i) \tag{11}$$

$$(\cos\theta)^{-1}\tilde{R}_j^i = \delta_j^i + \frac{\sin\theta}{\cos\theta}\tilde{L}_j^i + (\frac{1}{\cos\theta} - 1)(\tilde{L}_k^i \tilde{L}_j^k + \delta_j^i) \tag{12}$$

$$\tilde{R}_j^i = \delta_j^i + \sin\theta \cdot \tilde{L}_j^i + (1 - \cos\theta)\tilde{L}_k^i \tilde{L}_j^k \tag{13}$$

$$\tilde{L}_j^i = \frac{\cos\theta}{2\sin\theta}(u^i|_j - u^i|_j^T) \tag{14}$$

Hence, Chen Zhida's S-R decomposition can be extended to large deformation and large rotation cases with the means of discarding the unit-orthogonal rotation and replacing it with an orthogonal rotation with expanding or contracting. The paper shows that for large shear flow the correct definition of local average rotation angular should be defined by equation (9). For large shear, the average angular tends to $\frac{\pi}{2}$. If the local average rotation angular is more than $\frac{\pi}{2}$, then the continuity condition for continuum will be destroyed. So, we define the average angular range is $(-\frac{\pi}{2}, \frac{\pi}{2})$.

In the following research, the decomposition (2) will be referred as Chen's form-one and the decomposition (10) will be referred as Chen's form-two. The complex of fluid motion is that it has two intrinsic geometrical forms of motion modes. Later, it will become clear that the Chen's form-one corresponds to elastic flow and the Chen's form-two corresponds to irreversible flow.

## 3. Strain Rate Definition

For traditional mechanics of fluid motion, the strain rate is defined as:

$$S_j^i = \frac{1}{2}(u^i|_j + u^i|_j^T) \tag{15}$$

Considering the definition of local average rotation angular, the stretching strain rate should be defined as equation:

$$\tilde{S}_j^i = \frac{1}{2}(u^i|_j + u^i|_j^T) - (\frac{1}{\cos\theta} - 1)(\tilde{L}_k^i \tilde{L}_j^k + \delta_j^i) \quad \text{for irreversible flow} \tag{16}$$

$$S_j^i = \frac{1}{2}(u^i|_j + u^i|_j^T) - (1 - \cos\Theta)L_k^i L_j^k \quad \text{for elastic flow} \tag{17}$$

Analogy to Stokes' asymmetric strain rate, the rotation strain rate should be defined as equation:

$$\tilde{W}_j^i = (\cos\theta)^{-1}\tilde{R}_j^i - \delta_j^i = \frac{\sin\theta}{\cos\theta}\tilde{L}_j^i + (\frac{1}{\cos\theta} - 1)(\tilde{L}_k^i \tilde{L}_j^k + \delta_j^i) \quad \text{for irreversible flow} \tag{18}$$

$$W_j^i = R_j^i - \delta_j^i = \sin\Theta \cdot L_j^i + (1 - \cos\Theta)L_k^i L_j^k \quad \text{for elastic flow} \tag{19}$$

.

We can find that the coupling tensor $\tilde{C}_j^i = -(\frac{1}{\cos\theta} - 1)(L_k^i L_j^k + \delta_j^i)$, or $C_j^i = -(1 - \cos\Theta)L_k^i L_j^k$, which are completely determined by local average rotation feature, shows that for shearing fluid motion the traditional strain rate definition will significantly differ from the actual strain rate appeared in fluid.

To make the meaning more clear, we take the conventional rectangular coordinator system as the co-moving coordinator system defined on original configuration. In this case, we have [5]:

$$L^i_j = \begin{vmatrix} 0 & L_3 & -L_2 \\ -L_3 & 0 & L_1 \\ L_2 & -L_1 & 0 \end{vmatrix}, \quad (L_1)^2 + (L_2)^2 + (L_3)^2 = 1 \tag{20}$$

It is clear that $L_i$ are the components of unit vector along the local average rotation axis. For the average local rotation: $L_1 = L_2 = 0$, $L_3 = 1$, we fined that for Chen's form-two the coupling tensor has only one non-zero component: $\tilde{C}^3_3 = -(\frac{1}{\cos\theta} - 1)$. However, for Chen's form-one the coupling tensor has two non-zero components: $C^1_1 = C^2_2 = (1 - \cos\Theta)$. Hence, we can conclude that the stress determined by traditional strain rate is bigger than the intrinsic stress defined by Chen's form-two strain rate on the local average rotation axis direction and that the stress determined by traditional strain rate is less than the intrinsic stress defined by Chen's form-one strain rate on the local average rotation axis direction. It means that, no matter what stress level is, the non-symmetric parts of transformation (related with traditional vorticity) makes the two forms of decomposition be same possible. What will be the actual one? This problem only can be answered by the dynamics of fluid flow. Therefore, it must be historical. On the other hand, the coexistence may be possible for some fluid motion case. If this is true, only non-steady flow can meets it. Now, it is clear, the fluid motion is un-steady in essential sense. Hence, it is very important to define the local rotation related stress, or the stress related with vorticity in traditional sense. To this purpose, the rotation strain rate should be defined.

In the conventional rectangular coordinator system, according to the well known Navier-Stokes's wave equation, the stress definition for isotropic Newton fluid is:

$$\sigma^i_j = -p\delta^i_j + \lambda \frac{\partial u^l}{\partial x^l} \delta_{ij} + 2\mu \frac{\partial u^i}{\partial x^j} \tag{21}$$

where, $\lambda$ and $\mu$ are Lamé elastic constants of Newton fluid. It can be rewritten as:

$$\sigma^i_j = -p\delta^i_j + (\lambda \delta^i_j \delta^k_l + 2\mu \delta^i_l \delta^k_j)(S^l_k + W^l_k) \quad \text{for elastic flow} \tag{22}$$

$$\sigma^i_j = -p\delta^i_j + (\lambda \delta^i_j \delta^k_l + 2\mu \delta^i_l \delta^k_j)(\tilde{S}^l_k + \tilde{W}^l_k) \quad \text{for irreversible flow} \tag{23}$$

It means that for a given stress field, two flow modes may be produced. So, turbulent flow must be very common for Newton fluid. In non-linear fluid mechanics, it is studied by bifurcation theory. This research shows that the fluid motion has two intrinsic forms of geometric structure.

**4. Motion Equations for Complicated Motion of Fluid**

For fluids, it is meaningless to define the position of a particle as the particles are not only interchangeable but also have variant configurations. The continuum of fluid can not be simplified as the usual elastic continuum although the Lagrange coordinator system does mean that special material representation is under discussion. Mathematically, the material representation and be constructed by unit covering concept. If the number of particles within the unit volume covering is enough to present a significant macroscopic representation, it is reasonable to introduce macroscopic representation of the inertial velocity $u^i$ and density $\rho$ defined at the space point as:

$$u^i = \frac{1}{\sum_l m_l} \sum_k m_k V^i_k = \frac{1}{\rho} \sum_k m_k V^i_k, \quad \rho = \sum_l m_l \tag{24}$$

where, $m_j$ is the mass of particle "j", $V^l_j$ is its inertial velocity component. By this way, the fluid forms continuous field. As an artificially supposed assumption, all these macroscopic representation might be regarded as continuous and differentiable. For each unit volume covering in Lagrange coordinator system, the physical volume and its actual configuration are variant but its Lagrange coordinators are invariant.

For a unit covering material of fluid motion, its inner deformation energy can be expressed as:

$$W = W(F^i_j, \varphi), \quad W(\delta^i_j) = \varphi \tag{25}$$

By this energy definition, the total energy can be expressed as a polynomial:

$$W = \varphi + \frac{\partial W}{\partial F^i_j}(F^i_j - \delta^i_j) + \frac{\partial^2 W}{\partial F^i_j \partial F^k_l}(F^i_j - \delta^i_j)(F^k_l - \delta^k_l) + \cdots \tag{26}$$

Omitting higher order infinitesimals, for fluid motion the general form of stress is given by:

$$\sigma^i_j = \frac{\partial W}{\partial F^j_i} \tag{27}$$

Note that for no deformation case a static stress will exist as:

$$^0\sigma_j^i = \frac{\partial \varphi}{\partial \delta_i^j} = -\tilde{p}\delta_j^i \tag{28}$$

For solid material, as there is no particle exchange, this isotropic static stress does not exist except the effects of temperature are taken into consideration. It is this stress that makes the fluid has a flexible configuration in an unit covering.

For isotropic fluid material, the constitutive equation can be expressed as:

$$\sigma_j^i = -\tilde{p}\delta_j^i + \lambda(F_l^l - \delta_l^l)\delta_j^i + 2\mu(F_j^i - \delta_j^i) \tag{29}$$

Therefore, the above treatment is identical with traditional treatment. But, here, the stress is rationally defined. Based on above research, the definition of static pressure in conventional fluid mechanics is intrinsic.

For the instant commoving dragging coordinator system of unit covering, the material element motion equations are (Refer Appendix B):

$$\sigma_l^i\big|_l = \frac{\partial}{\partial t}(\rho u^i) + \rho u^l(F_l^i - \delta_l^i) \tag{30-1}$$

$$\sigma_j^i\big|_i = \frac{\partial}{\partial t}(\rho u^i F_j^i) + \rho u^i F_l^i(F_j^l - \delta_j^l) \tag{30-2}$$

$$e_{ijk}F_l^j\sigma_k^l = 0 \tag{30-3}$$

Intrinsically, the equations are highly non-linear.

For steady flow, one has:

$$\sigma_l^i\big|_l = \rho u^l(F_l^i - \delta_l^i) \tag{31-1}$$

$$\sigma_j^i\big|_i = \rho u^i F_l^i(F_j^l - \delta_j^l) \tag{31-2}$$

$$e_{ijk}F_l^j\sigma_k^l = 0 \tag{31-3}$$

For infinitesimal deformation, $F_j^i \approx \delta_j^i$, the equation (30) can be approximated as:

$$\sigma_l^i\big|_l = \frac{\partial}{\partial t}(\rho u^i) + \rho u^l(F_l^i - \delta_l^i), \quad e_{ijk}\sigma_k^j = 0 \tag{32}$$

It is the conventional NS equation. For this case, the stress must be symmetric. Hence, it can be inferred that for general fluid motion NS equation is not complete and hence cannot be closed.

Therefore, for fluid motion the conventional NS equation must be complemented by two additional equations. As these two equations are derived from the angular momentum conservation, a reasonable conclusion is that the conventional NS equation only meets linear momentum conservation and does not meet the angular momentum conservation.

Generally, for fluid motion, one has:

$$(\sigma_j^i - \sigma_j^{iT})\big|_i = \frac{\partial}{\partial t}[\rho u^i(F_j^i - \delta_j^i)] + \rho u^i[F_l^i(F_j^l - \delta_j^l) - \delta_l^i(F_j^l - \delta_j^l)^T] \tag{33}$$

Hence, for fluid motion the stress tensor is not symmetric.

On the other hand, one has:

$$(\sigma_j^i + \sigma_j^{iT})\big|_i = \frac{\partial}{\partial t}[\rho u^i(F_j^i + \delta_j^i)] + \rho u^i[F_l^i(F_j^l - \delta_j^l) + \delta_l^i(F_j^l - \delta_j^l)^T] \tag{34}$$

Introducing symmetric stress:

$$\pi_{ij} = \frac{1}{2}(\sigma_j^i + \sigma_j^{iT}) \tag{35}$$

The equation (34) can be expressed as:

$$\pi_{ij}\big|_i = \frac{\partial}{\partial t}(\rho u^i \delta_j^i) + \frac{1}{2}\rho u^l[(F_j^l - \delta_j^l) + (F_j^l - \delta_j^l)^T] + \delta\!f_j \tag{36}$$

where,

$$\delta\!f_j = \frac{1}{2}[\frac{\partial}{\partial t}(\rho u^i \cdot u^i\big|_j) + \rho u^i \cdot u^l\big|_j \cdot u^i\big|_l] \tag{37}$$

By this way, the strain rate is defined by:

$$s_{ij} = \frac{1}{2}[(F_j^l - \delta_j^l) + (F_j^l - \delta_j^l)^T] \tag{38}$$

And Reynolds stress $\theta_{ij}$ can be introduced by:

$$\delta\!f_j = \theta_{ij}\big|_i \tag{39}$$

Under above definition, the Reynolds fluid motion equation is derived as:

$$(\pi_{ij} - \theta_{ij})\big|_i = \frac{\partial}{\partial t}(\rho u^i \delta_j^i) + \rho u^l \cdot u^l\big|_j \tag{40}$$

Therefore, the Reynolds fluid motion equation is correct in form. However, in traditional treatment of statistic interpretation the Reynolds stress becomes an artificial item, which can be introduced by many possible ways and by different, even controversial, interpretation. This just is what happened in fluid mechanics.

Summering above discussion, both of the conventional NS fluid motion equation and the Reynolds fluid motion equation are included in the fluid motion equation (30). Therefore, the equation (30) will be used later for detailed study. For simplicity, the standard rectangular coordinator system will be taken as the commoving dragging coordinator system.

## 5. Incompressible Pure Rotation Flow

For incompressible fluid, the unit covering material will maintain its volume as an invariant. Chen's S+R decomposition form-one can be used for such a motion when $S_j^i = 0$. In this case, one will have:

$$\sigma_l^i\big|_l = \frac{\partial}{\partial t}(\rho u^i) + \rho u^l (R_l^i - \delta_l^i) \tag{41-1}$$

$$\sigma_j^i\big|_i = \frac{\partial}{\partial t}(\rho u^i R_j^i) + \rho u^i R_l^i (R_j^l - \delta_j^l) \tag{41-2}$$

$$e_{ijk} R_l^j \sigma_k^l = 0 \tag{41-3}$$

From equation (3), the condition $S_j^i = 0$ gives out:

$$s_{ij} = \frac{1}{2}(\frac{\partial u^i}{\partial x^j} + \frac{\partial u^j}{\partial x^i}) = (1 - \cos\Theta) L_k^i L_j^k \tag{42}$$

It shows that for a pure local rotation incompressible flow, the traditional strain rate $s_{ij}$ is determined by the local rotation axis direction and local average rotation angle $\Theta$. For pure rotation flow, the traditional strain rate is symmetric. This makes it cannot explain the local rotation motion. On the other hand, the local rotation in continuum is different from the rigid rotation in that it always require non-zero classical strain rate. From this point to see, the most suitable definition of pure rotation should be equation (42) or $S_j^i = 0$.

Notes that for plane rotation, if the rotation axis is along the $x^3$ direction, the equation (42) becomes:

$$s_{ij} = \frac{1}{2}(\frac{\partial u^i}{\partial x^j} + \frac{\partial u^j}{\partial x^i}) = (1 - \cos\Theta) L_k^i L_j^k = (1 - \cos\Theta) \begin{vmatrix} 1 & 0 & 0 \\ 0 & 1 & 0 \\ 0 & 0 & 0 \end{vmatrix} \tag{43}$$

That is to say that a pure plane local rotation will cause rotation plane symmetric velocity gradient. In this case, if one uses traditional strain rate definition to calculate stress, he will report that there is an additional plane pressure. As we know that such an additional plane pressure does not exist for pure local rotation flow, one may find that symmetric strain rate should be defined by $S_j^i$ rather than $s_{ij}$.

To see the $S_j^i = 0$ indeed defines the pure rotation motion, let us write down the velocity field variation in full form:

$$u^i = (\frac{\partial u^i}{\partial x^j} + \delta_j^i) U^j \tag{44}$$

where, $U^j$ represents the velocity of a neighboring reference space point.

When $S_j^i = 0$, one has: $\frac{\partial u^i}{\partial x^j} + \delta_j^i = R_j^i$. Hence:

$$u^i u^i = (R_k^i U^k)(R_l^i U^l) = R_k^i R_l^i U^k U^l = \delta_{kl} U^k U^l = U^k U^k \tag{45}$$

It shows that the velocity field is a pure rotation field. Most important feature of incompressible flow is that its kinetic energy is invariant. Therefore, it should be very comment phenomenon for incompressible fluid flow.

So, the pure rotation of fluid should be defined by $R_j^i$ for its rotation axis direction instead of the definition of vortex and by $\Theta$ for its magnitude instead of vortices.

For the pure rotation flow, the $L_j^i$ has only two independent parameters. So, equations (41) have six independent motion parameter, that is $u^i$, $L_j^i$, and $\Theta$, they form a complete motion equations. The equation (41-3) will degenerate in to an equation about pressure. So, the pressure is not intrinsic parameter of fluid, and it is motion-dependent. This equation is the fist one which gives analytical expression of fluid pressure.

To meet equation (41-3), the stress must be:
$$\sigma^i_j = 2\mu R^i_j \tag{46}$$
Since, by equation (29), one has:
$$\sigma^i_j = -\tilde{p}\delta^i_j + \lambda(R^l_l - \delta^l_l)\delta^i_j + 2\mu(R^i_j - \delta^i_j) \tag{47}$$
As $R^i_i - \delta^i_i = 2(1-\cos\Theta)$, it becomes:
$$\sigma^i_j = -\tilde{p}\delta^i_j + 2\lambda(1-\cos\Theta)\delta^i_j - 2\mu\delta^i_j + 2\mu R^i_j \tag{48}$$
To meet equation (41-3), the fluid motion must meet condition:
$$\tilde{p} = 2\lambda(1-\cos\Theta) - 2\mu \tag{49}$$

For incompressible pure rotation flow, the pressure and the local rotation angular are tightly related. How can this strong phenomenon happen? Viewing equation (41-1) and (41-2), the local rotation tensor is determinate. Hence, $(1-\cos\Theta)$ is determined. If the fluid feature parameters $\lambda$ and $\mu$ are invariant, the pressure $\tilde{p}$ must be variant. Therefore, the pressure $\tilde{p}$ is not intrinsic parameter of fluid. The equation (49) gives the calculation equation for the pressure $\tilde{p}$. If one supposes the pressure $\tilde{p}$ is constant, the fluid feature parameters $\lambda$ and $\mu$ are not intrinsic. They will be pressure-dependent or rotation-dependent. For rotational fluid flow, many controversial come from the definition of pressure. In most theoretic treatment, the pressure is artificial.

Based on this research, the equation (49) should be viewed as an equation which determines the fluid pressure.

Note that for infinitesimal rotation, the "static" pressure is: $\tilde{p} \approx -2\mu$. If $\mu$ is negative, the pressure is the normal positive pressure. However, if $\mu$ is positive, the pressure will be negative pressure. The turning point is determined by zero pressure state, that is:
$$\lambda(1-\cos\Theta_c) = \mu \tag{50}$$
Therefore, for a given fluid the incompressible pure rotation flow has a critical rotation angular at this local rotation angular the fluid pressure is zero. Mathematically, this point defines the instability of fluid motion.

## 6. Compressible Pure Rotational Flow

For incompressible fluid, the unit covering material will maintain its volume as an invariant. Chen's S+R decomposition form-two can be used for such a motion when $\tilde{S}^i_j = 0$. In this case, one will have:

$$\sigma^i_l\big|_l = \frac{\partial}{\partial t}(\rho u^i) + \rho u^l(\frac{1}{\cos\theta}\tilde{R}^i_l - \delta^i_l) \tag{51-1}$$

$$\sigma^i_j\big|_i = \frac{\partial}{\partial t}(\frac{1}{\cos\theta}\rho u^i \tilde{R}^i_j) + \frac{1}{\cos\theta}\rho u^i \tilde{R}^i_l(\frac{1}{\cos\theta}\tilde{R}^l_j - \delta^l_j) \tag{51-2}$$

$$e_{ijk}\tilde{R}^j_l\sigma^l_k = 0 \tag{51-3}$$

From equation (11), the condition $\tilde{S}^i_j = 0$ gives out the classical strain rate:
$$\tilde{s}_{ij} = \frac{1}{2}(u^i\big|_j + u^j\big|_i) = (\frac{1}{\cos\theta} - 1)(\tilde{L}^i_k\tilde{L}^k_j + \delta^i_j) \tag{52}$$
By the definition equation (14), introducing $\tilde{L}_i = e_{ijk}\tilde{L}^j_k$, the above equation can be rewritten as:
$$\tilde{s}_{ij} = (\frac{1}{\cos\theta} - 1)\tilde{L}_i\tilde{L}_j \tag{53}$$
It shows that for a pure local rotation compressible flow, the traditional strain rate $\tilde{s}_{ij}$ is determined by the local rotation axis direction and local average rotation angle $\theta$. For pure rotation flow, the traditional strain rate is symmetric. This makes it cannot explain the local rotation motion. On the other hand, the local rotation in continuum is different from the rigid rotation in that it always require non-zero classical strain rate. From this point to see, the most suitable definition of pure rotation of compressible fluid motion should be equation (53) or $\tilde{S}^i_j = 0$.

Notes that for plane rotation, if the rotation axis is along the $x^3$ direction, the equation (52) becomes:
$$\tilde{s}_{ij} = \frac{1}{2}(\frac{\partial u^i}{\partial x^j} + \frac{\partial u^j}{\partial x^i}) = (\frac{1}{\cos\theta} - 1)\begin{vmatrix} 0 & 0 & 0 \\ 0 & 0 & 0 \\ 0 & 0 & 1 \end{vmatrix} \tag{53}$$
Comparing it with the equation (43), the striking feature is that for pure rotation of compressible fluid motion,

only along the rotation axis direction the classical strain rate is non-zero, while for incompressible pure rotation only along the rotation axis direction the classical strain rate is zero. That is to say that a pure compressible local rotation will cause symmetric velocity gradient along rotation axis direction. In this case, if one uses traditional strain rate definition to calculate stress, he will report that there is an additional stress component along rotation axis.

To see the $\widetilde{S}_j^i = 0$ indeed defines the pure rotation motion, let us write down the velocity field variation in full form:

$$u^i = (\frac{\partial u^i}{\partial x^j} + \delta_j^i) U^j \tag{54}$$

where, $U^j$ represents the velocity of a initial reference space point.

When $\widetilde{S}_j^i = 0$, one has: $\frac{\partial u^i}{\partial x^j} + \delta_j^i = \frac{1}{\cos\theta} \widetilde{R}_j^i$. Hence:

$$u^i u^i = \frac{1}{\cos^2\theta}(\widetilde{R}_k^i U^k)(\widetilde{R}_l^i U^l) = \frac{1}{\cos^2\theta} U^k U^k \tag{55}$$

It shows that the velocity field is a pure increasing field. Most important feature of compressible flow is that its kinetic energy is increasing. Therefore, it should be very comment phenomenon for compressible fluid flow.

For the pure rotation flow, the $\widetilde{L}_j^i$ has only two independent parameters. So, equations (51) have six independent motion parameter, that is $u^i$, $\widetilde{L}_j^i$, and $\theta$, they form a complete motion equations. The equation (51-3) will degenerate in to an equation about pressure. So, the pressure is not intrinsic parameter of fluid, and it is motion-dependent. This equation is the fist one which gives analytical expression of fluid pressure.

To meet equation (51-3), the stress must be:

$$\sigma_j^i = 2\mu \frac{1}{\cos\theta} \widetilde{R}_j^i \tag{56}$$

Since, by equation (29), one has:

$$\sigma_j^i = -\widetilde{p}\delta_j^i + \lambda(\frac{1}{\cos\theta} \widetilde{R}_l^l - \delta_l^l)\delta_j^i + 2\mu(\frac{1}{\cos\theta} \widetilde{R}_j^i - \delta_j^i) \tag{57}$$

As $\frac{1}{\cos\theta}\widetilde{R}_i^i - \delta_i^i = \frac{1}{\cos\theta} - 1$, it becomes:

$$\sigma_j^i = -\widetilde{p}\delta_j^i + 2\lambda(\frac{1}{\cos\theta} - 1)\delta_j^i - 2\mu\delta_j^i + 2\mu\frac{1}{\cos\theta}\widetilde{R}_j^i \tag{58}$$

To meet equation (51-3), the fluid motion must meet condition:

$$\widetilde{p} = 2\lambda(\frac{1}{\cos\theta} - 1) - 2\mu \tag{59}$$

For compressible pure rotation flow, the pressure and the local rotation angular are tightly related. Viewing equation (51-1) and (51-2), the local rotation tensor is determinate. Hence, $(\frac{1}{\cos\theta} - 1)$ is determined. If the fluid feature parameters $\lambda$ and $\mu$ are invariant, the pressure $\widetilde{p}$ must be variant. Therefore, the pressure $\widetilde{p}$ is not intrinsic parameter of fluid. The equation (59) gives the calculation equation for the pressure $\widetilde{p}$. If one supposes the pressure $\widetilde{p}$ is constant, the fluid feature parameters $\lambda$ and $\mu$ are not intrinsic. They will be pressure-dependent or rotation-dependent. For rotational fluid flow, many controversial come from the definition of pressure. In most theoretic treatment, the pressure is artificial.

Based on this research, the equation (59) should be viewed as an equation which determines the fluid pressure for compressible rotation of fluid.

The turning point is determined by zero pressure state, that is:

$$\lambda(\frac{1}{\cos\theta_c} - 1) = \mu \tag{60}$$

Therefore, for a given fluid the compressible pure rotation flow has a critical rotation angular at this local rotation angular the fluid pressure is zero. Mathematically, this point defines the instability of fluid motion.

At the zero pressure state, both equations of (50) and (60) are true. Therefore, the instability point of fluid motion mode is determined by equation:

$$(\frac{1}{\cos\theta_c} - 1) = (1 - \cos\Theta_c) \tag{61}$$

In fact, for any kind of pure rotation, as the pressure $\widetilde{p}$ has two equation forms, the pure rotation motion is intrinsic changeable among Chen's form-one and Chen's form-two. That is to say the non-instability may appear if:

$$(\frac{1}{\cos\theta} - 1) = (1 - \cos\Theta) \tag{62}$$

For large rotation defined by:

$$\frac{1}{2}[(u^1|_2 - u^2|_1)^2 + (u^2|_3 - u^3|_2)^2 + (u^3|_1 - u^1|_3)^2]^{\frac{1}{2}} \geq 1 \tag{63}$$

The incompressible pure rotation mode cannot exist. In this case, only compressible pure rotation mode can exist. Therefore, large rotation is stable.

For compressible pure rotation mode, the gauge of unit covering is:

$$g_{ij} = \frac{1}{\cos^2\theta}\delta_{ij} \tag{64}$$

It means that it has volume expanding. Physically, the expanded volume is bubble volume density $\kappa$ for isothermal fluid motion. The bubble volume density $\kappa$ can be expressed as:

$$\kappa = \frac{1}{\cos^2\theta} - 1 \tag{65}$$

Then the conclusion can be made that the large rotation flow will produce bubble, whose volume density is completely determined by the local rotation angular.

## 7. Geometrical Structure of Turbulent Flow

For non-rotational flow, the motion equations are:

$$\sigma_l^i|_l = \frac{\partial}{\partial t}(\rho u^i) + \rho u^l S_l^i \tag{66-1}$$

$$\sigma_j^i|_i = \frac{\partial}{\partial t}[\rho u^i(S_j^i + \delta_j^i)] + \rho u^i S_l^i (S_j^l + \delta_j^l) \tag{66-2}$$

$$e_{ijk}(S_l^j + \delta_l^j)\sigma_k^l = 0 \tag{66-3}$$

In this case, it seems to be that the strain rate is symmetric and stress is symmetric, hence the equation (66-3) is met naturally. However, generally the stress is not symmetric. In fact, the equation (66-1) and (66-2) gives out:

$$(\sigma_j^i - \sigma_j^{iT})|_i = \frac{\partial}{\partial t}(\rho u^i S_j^i) + \rho u^i S_l^i S_j^l \tag{67}$$

So, only for steady infinitesimal flow defined by:

$$\frac{\partial}{\partial t}(\rho u^i S_i^j) \ll \frac{\partial}{\partial t}(\rho u^j) \tag{68-1}$$

$$\rho u^i S_l^i S_j^l \ll \rho u^i S_j^i \tag{68-2}$$

the stress can be approximated as symmetric. For steady infinitesimal flow, the motion equation can be approximated as:

$$\sigma_l^i|_l = \frac{\partial}{\partial t}(\rho u^i) + \rho u^l S_l^i \tag{69}$$

It is the traditional NS equation.

Therefore, the traditional NS equation is applicable only for steady infinitesimal flow.

For infinitesimal un-steady flow defined by equation (68-2), the motion equation can be approximated as:

$$\sigma_l^i|_l = \frac{\partial}{\partial t}(\rho u^i) + \rho u^l S_l^i \tag{70-1}$$

$$\sigma_j^i|_i = \frac{\partial}{\partial t}[\rho u^i(S_j^i + \delta_j^i)] + \rho u^i S_j^i \tag{70-2}$$

$$e_{ijk}(S_l^j + \delta_l^j)\sigma_k^l = 0 \tag{70-3}$$

For large velocity gradient flow, even the flow is steady absolutely that any time differential items are zero, the motion equation become:

$$\sigma_l^i|_l = \rho u^l S_l^i \tag{71-1}$$

$$\sigma_j^i|_i = \rho u^i S_l^i (S_j^l + \delta_j^l) \tag{71-2}$$

$$e_{ijk}(S_l^j + \delta_l^j)\sigma_k^l = 0 \tag{71-3}$$

The stress is still non-symmetric.

Summering up above discussion, it can conclude that: (a) large velocity gradient flow (no matter it is steady or non-steady) will produce non-symmetric stress; (b) un-steady flow (no matter it is infinitesimal or not) will produce non-symmetric stress.

If the non-symmetric stresses do exist, based on equation (29), there must be local rotation. That make the fluid flow must be rotational. If the local rotation is irreversible, accumulation effects will appear.

For the rotation mode defined by Chen's form-one, when the local rotation angular is big enough, it will change into the rotation mode defined by Chen's form-two. As the velocity will be keep increasing for compressible pure rotation flow, it can infer that the flow will become unstable.

In fact, at the critical point of instability, the local rotation will from conventional plane rotation change into axis expanding rotation. Then bubble will be produced. The flow will become complicated and un-steady. That is it becomes into turbulent flow.

Therefore, the geometrical structure of turbulent flow is:

$$\widetilde{F}_j^i = \widetilde{S}_j^i + (\cos\theta)^{-1} \widetilde{R}_j^i \tag{10}$$

Where:

$$\widetilde{S}_j^i = \frac{1}{2}(u^i\big|_j + u^i\big|_j^T) - (\frac{1}{\cos\theta} - 1)(\widetilde{L}_k^i \widetilde{L}_j^k + \delta_j^i) \tag{11}$$

$$(\cos\theta)^{-1} \widetilde{R}_j^i = \delta_j^i + \frac{\sin\theta}{\cos\theta}\widetilde{L}_j^i + (\frac{1}{\cos\theta} - 1)(\widetilde{L}_k^i \widetilde{L}_j^k + \delta_j^i) \tag{12}$$

$$\widetilde{R}_j^i = \delta_j^i + \sin\theta \cdot \widetilde{L}_j^i + (1-\cos\theta)\widetilde{L}_k^i \widetilde{L}_j^k \tag{13}$$

$$\widetilde{L}_j^i = \frac{\cos\theta}{2\sin\theta}(u^i\big|_j - u^i\big|_j^T) \tag{14}$$

As a comparing, before the flow changed into turbulent flow, the geometric structure of conventional flow is:

$$F_j^i = S_j^i + R_j^i \tag{2}$$

Where:

$$S_j^i = \frac{1}{2}(u^i\big|_j + u^i\big|_j^T) - (1-\cos\Theta)L_k^i L_j^k \tag{3}$$

$$R_j^i = \delta_j^i + \sin\Theta \cdot L_j^i + (1-\cos\Theta)L_k^i L_j^k \tag{4}$$

$$L_j^i = \frac{1}{2\sin\Theta}(u^i\big|_j - u^i\big|_j^T) \tag{5}$$

$$\sin\Theta = \frac{1}{2}[(u^1\big|_2 - u^2\big|_1)^2 + (u^2\big|_3 - u^3\big|_2)^2 + (u^3\big|_1 - u^1\big|_3)^2]^{\frac{1}{2}} \tag{6}$$

At the critical state, the deformation mode is not unique. From physical consideration, the stress field must be unique. As the stress is defined by equation (29), the critical condition by stress continuity is:

$$\frac{\sin\theta}{\cos\theta}\widetilde{L}_j^i = \sin\Theta \cdot L_j^i \tag{72}$$

Combing with the critical equation of rotation mode determined by strain rate continuity:

$$(\frac{1}{\cos\theta} - 1) = (1 - \cos\Theta) \tag{62}$$

the equations (72) and (62) form a complete description of turbulent transition.

If both of stress continuity and strain rate continuity are required, only limited and discrete $\theta$ and $\Theta$ can meet them. Therefore, turbulent flow is discrete in essential sense.

Once the flow becomes turbulent flow, bubble will be produced which is characterized by bubble volume density parameter $\kappa = \frac{1}{\cos^2\theta} - 1$ (65). Now, it is clear that the bubble will appear suddenly when the flow changed into turbulent flow.

Carefully examining equations (62), (65), and (72), it can conclude that the geometric structure of turbulent flow is completely determined by the initial conventional rotation flow which meets motion equation (31) with Chen's form-one decomposition.

Therefore, if using Chen's form decomposition in the fluid motion equation (31) with suitable boundary or initial conditions the parameters $L_j^i$ and $\Theta$ can be determined, then it is not turbulent flow. If the solution does not exist, then the flow must be turbulent flow. Then, by the upper limits of parameters $L_j^i$ and $\Theta$ for a practical problem, the turbulent transition condition can be estimated. Further, by the upper limits, the feature of turbulent flow can be calculated by equations (62) and (72). For many engineering problems, the bubble volume density can be calculated in this case.

In fact, the zero intrinsic stress corresponds to the instability of fluid motion. For this case, one has:

$$\lambda(1-\cos\Theta_c) = \mu \quad \text{for elastic flow} \tag{50}$$

$$\lambda(\frac{1}{\cos\theta_c}-1) = \mu \quad \text{for irreversible flow} \tag{60}$$

Therefore, for a given fluid material, the critical condition of turbulence can be determined by equations (50) or (60). It shows that the critical condition of turbulence is completely determined by the intrinsic parameters of fluid material.

Theoretically, the zero intrinsic stress state requires the pressure conditions:

$$\tilde{p} = 2\lambda(1-\cos\Theta) - 2\mu \quad \text{for elastic flow} \tag{49}$$

$$\tilde{p} = 2\lambda(\frac{1}{\cos\theta}-1) - 2\mu \quad \text{for irreversible flow} \tag{59}$$

Therefore, for a given pressure, the critical condition of turbulence can be determined by equations (49) or (59).

Summering above results, the research shows that there are three ways to determine the turbulent transition condition: (1) solving fluid motion equation; (2) using the intrinsic parameters of fluid material; (3) using pressure condition with given material parameters.

**8 Conclusion**

Traditionally, the incompressible flow is defined by:

$$\frac{\partial u^i}{\partial x^i} = 0 \tag{73}$$

Appling this equation to Chen's for-one decomposition, one will get:

$$R_i^i - 3 = 2(1-\cos\Theta) \tag{74}$$

Hence, the incompressible flow is defined by:

$$S_i^i = 2(1-\cos\Theta) \tag{75}$$

It corresponds to zero intrinsic stress discussed in section of incompressible pure rotation flow.

Based on this research, the incompressible flow can be divided into the addition of a pure rotation and a volume expanding caused by local average rotation. Hence, the traditional definition of impressible flow is correct when the local average rotation angle is small.

For incompressible Newton fluid, the traditional Navier-Stokes equation is:

$$\frac{\partial}{\partial t}(\rho u^i) = -\rho u^l (S_j^i + R_j^i - \delta_j^i) - \frac{\partial p_0}{\partial x^i} + \mu\frac{\partial}{\partial x^j}(S_j^i + R_j^i) - 2\lambda\frac{\partial\cos\Theta}{\partial x^i} \tag{76}$$

It can be seen that the local average rotation will produce an equivalent additional pressure field. The item ($2\lambda\frac{\partial\cos\Theta}{\partial x^i}$) is mist in the motion equation of traditional incompressible flow. So, it is understandable that the introduction of Reynolds stress will solve much more problems.

For compressible flow, one will get:

$$\frac{\partial u^i}{\partial x^i} = S_i^i + 2(1-\cos\Theta) \tag{77}$$

So, for compressible Newton fluid, the conventional Navier-Stokes equation is:

$$\frac{\partial}{\partial t}(\rho u^i) = -\rho u^l (S_j^i + R_j^i - \delta_j^i) - \frac{\partial p_0}{\partial x^i} + \mu\frac{\partial}{\partial x^j}(S_j^i + R_j^i) + \lambda\frac{\partial}{\partial x^i}[S_k^k + 2(1-\cos\Theta)] \tag{88}$$

For pure rotational flow ($S_j^i = 0$), the equation becomes:

$$\frac{\partial}{\partial t}(\rho u^i) = -\rho u^l (R_j^i - \delta_j^i) - \frac{\partial p_0}{\partial x^i} + \mu\frac{\partial}{\partial x^j}R_j^i - 2\lambda\frac{\partial\cos\Theta}{\partial x^i} \tag{89}$$

If one defines the "current pressure" $p$ as:

$$p = p_0 + 2\lambda\cos\Theta \tag{90}$$

Then equation (89) has an adjustable pressure $p$:

$$\frac{\partial}{\partial t}(\rho u^i) = -\rho u^l (R_j^i - \delta_j^i) - \frac{\partial p}{\partial x^i} + \mu\frac{\partial}{\partial x^j}R_j^i \tag{91}$$

Therefore, for pure rotation flow of Newton fluid, as equation (89) holds, there still exists an additional pressure increment [8]. This explains why in the calculation of rotational flow the pressure becomes an adjustable parameter.

Finally, it worth to point out that the ($2\lambda\cos\Theta\cdot\delta_j^i$) item play the similar role of Reynolds stress in form. However, this research does not support to introduce the Reynolds stress by the statistical way for Newton fluid.

The research presented here is focused on the physical foundation of fluid motion equation. By introducing the Chen's S+R decomposition, the research shows that for Newton fluid, whether it is incompressible or compressible in traditional definition, the local average rotation always produces an additional pressure equivalent item in the traditional Navier-Stokes equation. This item is deterministic rather than stochastic (as Reynolds stress) or adjustable (as exposed by this paper). The research have supplied a solvable mathematical model for the calculation of velocity and pressure when suitable boundary condition is adapted.

The research shows that the Navier-Stokes equation is good approximation for average flow, so it can not solve the turbulent problem in essential sense. However, this conclusion does not deny that with suitable additional condition for special engineering problem it is still a would-be acceptable approximation. The research shows that the classical Navier-Stokes equation does not meet the requirement of angular momentum conservation, which is apparently ignored for infinitesimal velocity gradient of fluid. The research concludes that: (a) large velocity gradient flow (no matter it is steady or non-steady) will produce non-symmetric stress; (b) un-steady flow (no matter it is infinitesimal or not) will produce non-symmetric stress. The non-symmetric stress produces local rotation. When the local rotation is big enough, the flow will become turbulent flow. The research shows that there are three ways to determine the turbulent transition condition: (1) solving fluid motion equation; (2) using the intrinsic parameters of fluid material; (3) using pressure condition with given material parameters. Therefore, the essential feature turbulent flow is well described in this research.

At present stage, the related results are only theoretic. The further experimental research is waiting to be done.

**References**

[1]Charles L. Fefferman, Existence & smoothness of the Navier-Stokes equation, Princeton Univ. Department of Mathematics, Princeton, N J 08544-1000, 2000
[2]Yang Benluo, Logic self-consistency analyses of theoretical fluid dynamics — philosophical and mathematical thinking originated from turbulence flow, Shanghai Jiao Tong University Press, (in Chinese) 1998, 514
[3]Yang Benluo, Turbulence and reconstruction of theoretical fluid dynamics, Shanghai JiaoTong University Press, (in Chinese) 2003, 351
[4]Lodge A S. Body tensor fields in continuum mechanics. Academic Press, 1974
[5]Chen Zhida. Rational Mechanics—Non-linear Mechanics of Continuum. Xizhou: China University of Mining & Technology Press, (In Chinese) 1987, 308
[6]Xiao Jianhua, Decomposition of displacement gradient and strain definition, in Y Luo, R Qiu ed. *Advance in Rheology and its Application (2005)*, Science Press USA Inc. 2005, 864-868
[7]Claeyssen, J.R., R.B. Plate and E. Bravo, Simulation in primitive variables of incompressible flow with pressure Neumann condition. Int. J. Mech. Fluids 30: 1009-1026, 1999
[8]Lien, F.S., A pressure-based unstructured grid method for all-speed flows. Int. J. Numer. Meth. Fluids 33: 355-374, 2000
[9]Chen Zhida, Rational Mechanics, Chongqing: Chongqing Pub., 2000 (in Chinese)
[10]Noda, H., A. Nakayama, Free-stream turbulence effects on the instantaneous pressure and forces on cylinder of rectangular cross section, Experiments in Fluids 34, 332-344, 2003

**Appendix A    Mathematic Feature of Deformation Tensor**

For continuum, each material point can be parameterized with continuous coordinators $x^i, i=1,2,3$. When the coordinators are fixed for each material point no matter what motion or deformation happens, the covariant gauge field $g_{ij}$ at time $t$ will define the configuration in the time. The continuous coordinators endowed with the gauge field tensor define a co-moving dragging coordinator system.

The initial configuration gauge $g_{ij}^0$ defines a distance geometric invariant:

$$ds_0^2 = g_{ij}^0 dx^i dx^j \tag{A-1}$$

The symmetry and positive feature of gauge tensor ensures that there exist three initial covariant base vectors $\vec{g}_i^0$ make:

$$g_{ij}^0 = \vec{g}_i^0 \cdot \vec{g}_j^0 \tag{A-2}$$

For current configuration, three current covariant base vectors $\vec{g}_i$ exist which make:

$$g_{ij} = \vec{g}_i \cdot \vec{g}_j \tag{A-3}$$

The current distance geometric invariant is:

$$ds^2 = g_{ij} dx^i dx^j \tag{A-4}$$

For each material point, there exists a local transformation $F_j^i$, which relates the current covariant base vectors with initial covariant base vectors:

$$\vec{g}_i = F_i^j \vec{g}_j^0 \tag{A-5}$$

So, the current covariant gauge tensor can be expressed as:

$$g_{ij} = F_i^k F_j^l g_{kl}^0 \tag{A-6}$$

In Riemann geometry, contra-variant gauge tensor $g^{ij}$, $g^{0ij}$ can be introduced, which meets condition:

$$g^{il} g_{jl} = \delta_j^i, \quad g^{0il} g_{jl}^0 = \delta_j^i \tag{A-7}$$

Similarly, contra-variant base vectors $\vec{g}^i$, $\vec{g}^{0i}$ can be introduced for current configuration and initial configuration respectively. Mathematically, there are:

$$g^{ij} = \vec{g}^i \cdot \vec{g}^j, \quad g^{0ij} = \vec{g}^{0i} \cdot \vec{g}^{0j} \tag{A-8}$$

There exists a local transformation $G_j^i$, which relates the contra-variant base vectors:

$$\vec{g}^i = G_j^i \vec{g}^{0j} \tag{A-9}$$

So, the current contra-variant gauge tensor can be expressed as:

$$g^{ij} = G_k^i G_l^j g^{0kl} \tag{A-10}$$

By equations (A-6), (A-7), and (A-10), it is easy to find out that:

$$G_l^i F_j^l = \delta_j^i \tag{A-11}$$

Hence, the transformation $F_j^i$ relates the initial contra-variant base vectors with current contra-variant base vectors in such a way that:

$$\vec{g}^{0i} = F_j^i \vec{g}^j \tag{A-12}$$

Therefore, the transformation $F_j^i$ is a mixture tensor. Its lower index represents covariant component in $g_{ij}^0$ configuration, its upper index represents contra-variant component in $g_{ij}^0$ configuration.

Similar discussion shows that the local transformation $G_j^i$ is a mixture tensor, lower index represents covariant component in $g_{ij}$ configuration, upper index represents contra-variant component in $g_{ij}$ configuration. It is easy to find that:

$$\vec{g}_i^0 = G_i^j \vec{g}_j \tag{A-13}$$

Other two important equations are:

$$F_j^i = \vec{g}^{0i} \cdot \vec{g}_j \tag{A-14}$$

$$G_j^i = \vec{g}^i \cdot \vec{g}_j^0 \tag{A-15}$$

By these equations, $F_j^i$ can be explained as the extended-Kronecker-delta in that it's the dot product of contra-variant base vector in initial configuration and covariant base vector in current configuration. When the two configurations are the same, it becomes the standard Kronecker-delta. For the $G_j^i$, similar interpretation can be made.

From these equations to see, geometrically, it is found that the $F_j^i$ measures the current covariant base vector with the initial contra-variant base vector as reference. When stress tensor is defined as:

$$\sigma_j^i = E_{jl}^{ik}(F_k^l - \delta_k^l) \tag{A-16}$$

Its mechanic meaning can be explained as the lower index represents component of surface force in current direction, and the upper index represents the initial surface normal which surface force acts on. It takes the initial surface as surface force reference.

The $G_j^i$ measures the initial covariant base vector with the current contra-variant base vector as reference. The corresponding stress can be explained as the lower index represents component of surface force in initial direction, and the upper index represents the current surface normal which surface force acts on.

For such a kind of mixture tensor, it is defined on the same point with different configurations (that is initial and current). So, the name of two-point tensor given by C Truessdell is not correct. This mis-interpretation has caused many doubts cast on the feature of transformation tensor. Some mathematician even said that the mixture tensor is meaningless.

However, during treating the non-symmetric field theory, Einstein believes that the use of mixture tensor is more reasonable that the pure covariant or pure contra-variant tensor. Recently, the concept of bi-tensor is introduced in physics topic. So, we have sound reason to use mixture tensor in continuum mechanics, as it can give clear physical meaning for the definition of stress.

Note that the transformation $F_j^i$ is completely determined by the deformation measured in initial configuration with gauge tensor $g_{ij}^0$. Mathematically, the covariant differentiation is taken in the initial geometry also, although the physical meaning of $F_j^i$ is that it relates two configurations.

It is valuable to point out that, if the initial coordinator system is taken as Cartesian system, the $F_j^i$ can be transformed into pure covariant form:

$$F_{ij} = \vec{g}_i^{\,0} \cdot \vec{g}_j \tag{A-17}$$

The $G_j^i$ can be transformed into pure contra-variant form:

$$G^{ij} = \vec{g}^{\,i} \cdot \vec{g}^{\,0j} \tag{A-18}$$

Although it is acceptable in form for the special case of taking Cartesian system as the initial coordinator system, the intrinsic meaning of deformation tensor is completely destroyed by such a formulation. That may be the main reason for the rational mechanics constructed by C Truessdell et al in 1960s.

Mathematically, once the initial gauge field is selected, the current gauge field is to be obtained by the given physical deformation. In this sense, the current gauge field is viewed as the physical field. So, the covariant differentiation is taken respect with the initial configuration.

Truessdell argued that the covariant differentiation should be taken one index in initial configuration, another index in current configuration. This concept had strongly effects on the development of infinite deformation mechanics. Such a kind of misunderstanding indeed is caused by the equations (A17-18).

Historically, Chen Zhida is the first one to clear the ambiguity caused by equations (A-17) and (A-18) systematically. His monograph "Rational Mechanics"(1987) treats this topic in depth.

There are many critics about the mathematics used in Chen's rational mechanics theory. The most common one is that: for a coordinator transformation

$$dx^i = A_j^i dX^j, \quad dX^i = (A_j^i)^{-1} dx^j \tag{A-19}$$

the covariant and contra-variant components of tensor is defined by its transformation from new coordinator system to old coordinator system follows $A_j^i$ or $(A_j^i)^{-1}$ formulations. However, such a tensor definition is to make:

$$ds^2 = g_{ij} dx^i dx^j = G_{ij} dX^i dX^j \tag{A-20}$$

be invariant. Such a tensor describes a continuum without any deformation. In fact, such a tensor feature is only to say the object indifference for coordinator system selection.

However, many physicists and mechanists treat motion of continuum as the transformation $A_j^i$ or $(A_j^i)^{-1}$. Mathematically, the equation (A-20) can be rewritten as:

$$ds^2 = g_{ij} dx^i dx^j = g_{kl} A_i^k A_j^l dX^i dX^j = G_{ij} dX^i dX^j \tag{A-21}$$

or:

$$ds^2 = G_{ij} dX^i dX^j = G_{kl} (A_i^k)^{-1} (A_j^l)^{-1} dx^i dx^j = g_{ij} dx^i dx^j \tag{A-22}$$

It is clear that the covariant invariant feature must be maintained by the coordinator system choice. It has no any meaning of motion. In Chen's geometry, for time parameter $t$,

$$ds^2(t) = g_{ij}(t) dx^i dx^j = g_{kl}(0) F_i^k(t) F_j^l(t) dx^i dx^j \tag{A-23}$$

The gauge field is time dependent, while the coordinator is fixed (called intrinsic coordinator).

The equation (A-23) can be rewritten as:

$$ds^2(t) = g_{ij}(t) dx^i dx^j = g_{kl}(0) F_i^k(t) F_j^l(t) dx^i dx^j = g_{kl}(0) dX^k(t) dX^l(t) \tag{A-24}$$

It is similar in form with equation (A-21). But their mechanical interpretation is strikingly different.

For fluid motion, the transformation of base vector is determined by the inertial velocity gradient. In this case, the current base vector is time-dependent.

**Appendix B: Motion Equation of Fluid Flow**

For the displacement field defined in the initial co-moving dragging coordinator system, the displacement or rotation of continuum as a rigid whole body has no contribution to the intrinsic deformation. So we only consider the conservation related with deformation. In the co-moving dragging coordinator system, the linear momentum conservation can be expressed in integral form as:

$$\frac{\partial}{\partial t}\int_\Omega (\rho u^i)d\Omega = \oint_S \sigma_l^i n^l da \tag{B-1}$$

where, $n^l$ represent the unit normal vector of outside surface, $\rho$ is mass density at current configuration, $d\Omega$, $da$ represent the integration is taken at current configuration.

Based on Stokes Law, it is easy to find the differential form of linear momentum conservation:

$$(\sigma_l^i)\big|_l = \frac{\partial}{\partial t}(\rho u^i) \tag{B-2}$$

In the co-moving dragging coordinator system, the angular momentum conservation can be expressed in integral form as:

$$\frac{\partial}{\partial t}\int_\Omega (\vec{r}+\delta\vec{u})\times(\rho\vec{u})d\Omega = \oint_S (\vec{r}+\delta\vec{u})\times \vec{t}\,da \tag{B-3}$$

where, $\vec{r}$ is the position vector of material element in co-moving dragging coordinator system, $\delta\vec{u}$ is local velocity difference vector, whose geometric meaning is the displacement of unit time, $t_j = \sigma_j^l n_l$, $n_l$ is the unit tangent vector of outside surface.

Note that $F_j^i = u^i\big|_j + \delta_j^i$, introducing extended Kronecker $e_{ijk}$, the equation can be expressed as:

$$\int_\Omega \sqrt{g^0}\,e_{ijk}(x^j+\delta u^j)\frac{\partial}{\partial t}(\rho u^k)d\Omega = \int_\Omega \sqrt{g^0}\,e_{ijk}(x^j+\delta u^j)\sigma_k^m\big|_m d\Omega \tag{B-4}$$

that is:

$$\int_\Omega \sqrt{g^0}\,e_{ijk}(x^j+\delta u^j)\frac{\partial}{\partial t}(\rho u^k)d\Omega = \int_\Omega \sqrt{g^0}\,e_{ijk}F_m^j \sigma_k^m d\Omega + \int_\Omega \sqrt{g^0}\,e_{ijk}(x^j+\delta u^j)(\sigma_k^m)\big|_m d\Omega \tag{B-5}$$

where, $g^0 = \det\big|g_{ij}^0\big|$. Note that the velocity in initial configuration can be converted into velocity in current configuration as bellow:

$$u^i \vec{g}_i^0 = u^i g_{ij}^0 \vec{g}^{0j} = u^i g_{ij}^0 F_k^j \vec{g}^k \tag{B-6}$$

Hence, the differential form of angular momentum conservation is:

$$(\sigma_j^i)\big|_i = \frac{\partial}{\partial t}(\rho u^i g_{il}^0 F_j^l) \tag{B-7}$$

$$e_{ijk}F_l^j \sigma_k^l = 0 \tag{B-8}$$

Therefore, the motion equation of deformation expressed by displacement field $U^i$ is:

$$(\sigma_l^i)\big|_l = \frac{\partial}{\partial t}(\rho u^i) \tag{B-9-1}$$

$$(\sigma_j^i)\big|_i = \frac{\partial}{\partial t}(\rho u^i g_{il}^0 F_j^l) \tag{B-9-2}$$

$$e_{ijk}F_l^j \sigma_k^l = 0 \tag{B-9-3}$$

For large deformation, as the lower index of stress represents its component in current configuration, so the equation (B-9-2) is related with the velocity in current configuration.

If local average rotation is not considered, that is for the deformation:

$$F_j^i = S_j^i + \delta_j^i \tag{B-10}$$

The equation (B-9-3) will be met by a symmetric stress $\sigma_j^i$. If the stress $\sigma_j^i$ is symmetric, the equation (B-9-3) will require that the deformation $F_j^i$ must be symmetric. However, this can be true only for steady fluid motion. Where, the steady motion equations are:

$$(\sigma_l^i)\big|_l = 0 \tag{B-10-1}$$

$$(\sigma_j^i)\big|_i = 0 \tag{B-10-2}$$

$$e_{ijk}F_l^j \sigma_k^l = 0 \tag{B-10-3}$$

So, even for large deformation, the steady stress must be symmetric, and hence, the steady deformation must be symmetric. The conclusion is that the traditional fluid dynamics theory is correct. However, for dynamic fluid motion, it is correct only for infinitesimal deformation.

In classical infinitesimal deformation fluid motion, $F_j^i \approx \delta_j^i$, so the motion equation (B-9) can be

approximated as:

$$\frac{\partial \sigma_{ij}}{\partial x^j} = \frac{\partial}{\partial t}(\rho u^i) \tag{B-11-1}$$

$$\sigma_{ij} = \sigma_{ji} = -p\delta_{ij} + C_{ij}^{kl}\varepsilon_{kl} \tag{B-11-2}$$

$$\varepsilon_{ij} = \frac{1}{2}(\frac{\partial u^i}{\partial x^j} + \frac{\partial u^j}{\partial x^i}) \tag{B-11-3}$$

In fact, the equations form the bases of traditional Newton fluid motion theory.

For fluid material motion, instant commoving dragging coordinator system can be taken to define the instant fluid material. In this case, the inertial coordinator can be used to define the initial commoving dragging coordinator system. By this definition, the rate of base vector has contribution to the Lagrange time differentiation. For simplicity, the inertial coordinator system can be selected as standard rectangular.

For such a dynamic base vector, one has[5]:

$$\frac{\partial}{\partial t}(u^i \vec{g}_i) = (\frac{\partial u^i}{\partial t} + u^j \cdot u^i\big|_j)\vec{g}_i \tag{B-12}$$

Then, the motion equation should be rewritten as:

$$\sigma_l^i\big|_l = \frac{\partial}{\partial t}(\rho u^i) + \rho u^l(F_l^i - \delta_l^i) \tag{B-13-1}$$

$$\sigma_j^i\big|_i = \frac{\partial}{\partial t}(\rho u^i F_j^i) + \rho u^i F_l^i(F_j^l - \delta_j^l) \tag{B-13-2}$$

$$e_{ijk}F_l^j\sigma_k^l = 0 \tag{B-13-3}$$

Where, Eddington's general dimension principle[5] is used.

Correspondingly, when the deformation rate of fluid is taken into consideration, the equation (B-11) becomes:

$$\frac{\partial \sigma_{ij}}{\partial x^j} = \frac{\partial}{\partial t}(\rho u^i) + \rho u^j \cdot \frac{\partial u^i}{\partial x^j} \tag{B-14-1}$$

$$\sigma_{ij} = \sigma_{ji} = -p\delta_{ij} + C_{ij}^{kl}\varepsilon_{kl} \tag{B-14-2}$$

$$\varepsilon_{ij} = \frac{1}{2}(\frac{\partial u^i}{\partial x^j} + \frac{\partial u^j}{\partial x^i}) \tag{B-14-3}$$

It just is the usual NS equation for fluid motion.